\documentstyle[12pt]{article}

\newcommand{\dslash}{\partial\!\!\!/}
\newcommand{\aslash}{A\!\!\!/}
\newcommand{\Dslash}{D\!\!\!\!/}
\begin{document}
\thispagestyle{empty}
\begin{raggedleft}
hep-th/9707204\\
\end{raggedleft}
$\phantom{x}$\vskip 0.618cm\par
{\huge \begin{center}BOSE SYMMETRY AND CHIRAL 
DECOMPOSITION OF 2D FERMIONIC DETERMINANTS
\footnote{This work is 
supported by CNPq, Bras\'\i lia, Brasil}
\end{center}}\par

\begin{center}
$\phantom{X}$\\
{\Large E.M.C.Abreu, R.Banerjee\footnote{On leave
of absence from S.N.Bose National Centre for Basic
Sciences, Calcutta, India. e-mail:rabin@if.ufrj.br} and C.Wotzasek}\\[3ex]
{\em Instituto de F\'\i sica\\
Universidade Federal do Rio de Janeiro\\
21945, Rio de Janeiro, Brazil\\}
\end{center}\par
\begin{abstract}

\noindent We show in a precise way, either in the fermionic or its bosonized
 version, that Bose symmetry provides a systematic way to carry out the chiral
 decomposition of the two dimensional fermionic determinant.  
Interpreted properly, we show that there is no obstruction of this
 decomposition to gauge
invariance, as is usually claimed. Finally,
a new way of interpreting the Polyakov-Wiegman identity is proposed.
\end{abstract}
\vfill
\newpage

It is often claimed\cite{AAR} that the chiral 
decomposition of the two dimensional 
fermionic determinant poses an obstruction to gauge invariance.  
In this paper we clarify several aspects of this decomposition.  Contrary 
to the usual approach, the inverse route, whereby two chiral components 
are fused or soldered, is also examined in details. A close correspondence 
between the splitting and the soldering processes is established.  
By following Bose symmetry it is possible to give explicit expressions for 
the chiral determinants which show, in both these procedures, that there is
no incompatibility with gauge invariance at the quantum level.
Two important consequences emerging from this analysis are the close
connection 
between Bose symmetry and gauge invariance, 
and a novel interpretation of the Polyakov-Wiegman identity\cite{PW}.  

It is worth mentioning that understanding the properties of
2D-fermionic determinats and the associated role of Bose symmetry is crucial 
because of several aspects. For instance, the precise form of the one cocycle 
necessary in the recent discussions on smooth functional bosonisation
\cite{DN, DN1} is only
dictated by this symmetry \cite{RB}. Furthermore this cocycle, which
is just the 2D anomaly,
is known to be the origin of anomalies in higher dimensions by a set
of descent equations
\cite{RJ}. Incidentally, the anomaly phenomenon still defies a complete
explanation.

To briefly recapitulate the problem of chiral decomposition,
 consider the vacuum functional,
\begin{eqnarray}
\label{10}
e^{iW[A]} &=& \int d\bar\psi \; d\psi \exp\{i \int d^2 x \:
\bar\psi(i \dslash + e \aslash )\psi\}\nonumber\\
&=&\det(i\dslash +e \aslash)
\end{eqnarray}

\noindent where the expression for the determinant follows immediately by
imposing gauge invariance,

\begin{equation}
\label{20}
W[A] = N\int d^2x\: A_\mu \Pi^{\mu\nu}A_\nu
\end{equation}

\noindent with $\Pi^{\mu\nu}=g^{\mu\nu} - 
{\partial^\mu\partial^\nu\over \Box}\;\; ;\; \mu ,\nu=0,1\;$ 
being the transverse projector.  An explicit one loop calculation 
yields \cite{JS} $N={e^2\over 2\pi}\;$. Introducing light-cone variables,

\begin{equation}
\label{30}
A_\pm = {1\over\sqrt 2}(A_0\pm A_1)=A^\mp \;\;\; ;\;\;\; \partial_\pm=
{1\over\sqrt 2}(\partial_0\pm \partial_1)=\partial^\mp
\end{equation}

\noindent with the projector matrix given by,

\begin{equation}
\label{40}
\Pi^{\mu\nu}= {1\over 2}\left( 
\begin{array}{cc}
{-{\partial_-\over \partial_+}} & 1 \\
1 & {-{\partial_-\over \partial_+}} 
\end{array}
\right)
\end{equation}

\noindent it is simple to rewrite (\ref{20}) as,

\begin{equation}
\label{50}
W[A_+,A_-]=  -{ N\over 2} \int d^2x\: \{A_+ {\partial_-\over \partial_+}A_+ 
+ A_- {\partial_+\over \partial_-}A_- - 2 A_+ A_-\}
\end{equation} 

\noindent The factorization of (\ref{10}) into its chiral components yields,

\begin{equation}
\label{60}
\det(i\dslash +e \aslash)=\det(i\dslash +e \aslash_+)\det(i\dslash +e 
\aslash_-)
\end{equation}

\noindent where ${\aslash}_\pm ={\aslash} P_{\pm}$ with the chiral
projector defined
as $P_\pm ={1\pm\gamma_5\over 2}$. 
The effective action for the vector theory in terms of the chiral
components is now obtained from (\ref{60}), leading to an effective action,

\begin{eqnarray}
\label{70}
W_{eff}&=&  -{ N\over 2} \int d^2x\: \{A_+ {\partial_-\over 
\partial_+}A_+ + A_- {\partial_+\over \partial_-}A_- \}\nonumber\\
& =& W[A_+,0]+W[0,A_-]
\end{eqnarray}

\noindent which does not reproduce the expected gauge invariant
result (\ref{50}).  The above factorization is therefore regarded
as an obstruction to gauge invariance.

It is important to notice that (\ref{70}) follows from (\ref{60}) only
if one naively computes the chiral determinants from the usual vector
case (\ref{50}) by substituting either $A_+=0$ or $A_-=0$.  This may
be expected naturally since $\det(i\dslash +e \aslash)=\det(i\dslash 
+e \aslash_+ + e \aslash_-)$.  But the point is that whereas the usual
Dirac operator has a well defined eigenvalue problem,

\begin{equation}
\label{80}
\Dslash \:\psi_n =(i\dslash + e\aslash)\psi_n=\lambda_n\psi_n
\end{equation}

\noindent with the determinant being defined 
by the product of its eigenvalues, this is not true for the chiral pieces 
in the RHS of (\ref{60}), which lacks a definite eigenvalue equation \cite
{GW, RB} because the kernels map from one chiral sector to the other,

\begin{equation}
\label{90}
\Dslash_\pm \psi_\pm = \lambda\psi_\mp
\end{equation}

\noindent with $\psi_\pm = P_\pm\psi $.  Consequently, it is not possible
to interpret, however loosely or naively, any expression obtainable from
$\det\Dslash$ by setting $A_\pm=0$, as characterising $\det{\Dslash}_\mp$.

Since $\det{\Dslash}_\pm$ are not to be
regarded as $W[A_+,0]$ or $W[0,A_-]$ in (\ref{70}),
it is instructive to clarify the meaning of the latter expressions. 
Reconsidering $\det\Dslash$ as $\det(i\dslash +e \aslash_+ + e \aslash_-)$
it is easy 
to observe that the fundamental fermion loop decomposes into four 
pieces {\bf(see figure)}.

At the unregularized level there are different choices of 
interpreting these diagrams, depending on the location of the chiral 
projectors $P_\pm$.  In particular, by pushing one of these projectors 
through the loop and inserting it at the other vertex would yield vanishing 
contributions for the last two diagrams, since $P_+ P_- =0$.  It was shown 
earlier by one of us \cite{RB}, in a different context, that Bose symmetry
provided a 
definite guideline in manipulating such diagrams.  In other words, the 
position of the projectors is to be preserved exactly as appearing above, 
and the contributions explicitly computed from (\ref{20}) by appropriate 
replacements at the vertices.  This procedure implies a consistent way of
regularizing all four graphs.  Thus,

\begin{eqnarray}
\label{100}
W_{1(2)}&=& N\int d^2x\: A_\mu {\cal P}^{\mu\nu}_{+(-)}\Pi_{\nu\alpha}
{\cal P}^{\alpha\beta}_{-(+)}A_\beta\nonumber\\
W_{3(4)}&=& N\int d^2x\: A_\mu {\cal P}^{\mu\nu}_{+(-)}\Pi_{\nu\alpha}
{\cal P}^{\alpha\beta}_{+(-)}A_\beta
\end{eqnarray}

\noindent where
   
\begin{equation}
\label{110}
{\cal P}^{\alpha\beta}_{+(-)}={1\over 2}
(g^{\alpha\beta}\pm\epsilon^{\alpha\beta}) \;\;\; ;\epsilon^{+-}=
\epsilon_{-+}=1
\end{equation}

\noindent Using (\ref{40}) it is easy to simplify (\ref{100}) as,

\begin{equation}
\label{120}
W_1=W[A_+,0]\;\; ,\;\; W_2=W[0,A_-]\;\; ,\;\; W_3=W_4={N\over 2}A_+A_-
\end{equation}

\noindent Adding all four terms exactly reproduces the
gauge invariant result (\ref{50}).  If, on the contrary,
Bose symmetry was spoilt in the last two graphs as indicated
earlier so that $W_3=W_4=0$, the gauge noninvariant structure
(\ref{70}) is obtained.  This shows the close connection between
Bose symmetry and gauge invariance.  Recall that the same is also
true in obtaining the ABJ anomaly from the triangle graph\cite{R, SA}. 
Furthermore $W[A_+,0]$ and $W[0,A_-]$ are now seen to correspond to
graphs $W_1$ and $W_2$, respectively, evaluated in a very specific
fashion.  It is also evident that the incorrect manner of abstracting
$\det(i\dslash +e \aslash_{\pm})$ from $\det(i\dslash +e \aslash_)$
violates Bose symmetry leading to an apparent contradiction between
chiral factorization and gauge invariance.  Consequently the possibility
of ironing out this contradiction exists by interpreting the chiral
determinants as,

\begin{eqnarray}
\label{130}
-i\ln \det(i\dslash +e \aslash_+)&=& W[A_+,0]+{N\over 2}\int d^2x\: A_+ A_-
\nonumber\\
-i\ln \det(i\dslash +e \aslash_-)&=& W[0,A_-] +{N\over 2}\int d^2x\: A_+ A_-
\end{eqnarray}

\noindent These expressions just reduce to the naive definitions if
the crossing graphs are ignored or, equivalently, Bose symmetry is violated.

To put (\ref{130}) on a solid basis it must be recalled that (\ref{60}), 
as it stands, is only a formal identity.  A definite meaning can be attached 
provided some regularization is invoked to explicitly define the determinants
 appearing on either side of the equation.  Using a regularization that 
preserves the vector gauge symmetry of the LHS of (\ref{60}) led to the 
expression (\ref{50}).  As is well known \cite{RJ, JR} there is no 
regularization that retains the chiral symmetry of the pieces in the 
RHS of (\ref{60}).  An explicit one loop computation yields \cite{JR, RB1}, 
in a bosonized language,

\begin{eqnarray}
\label{140}
W_+[\varphi] &=& {1\over{4\pi}}\int d^2x\,\left(\partial_+
\varphi\partial_-\varphi +2 \, e\,A_+\partial_-\varphi + a\, 
e^2\, A_+ A_-\right)\nonumber\\
W_-[\rho]&=& {1\over{4\pi}}\int d^2x\,\left(\partial_+\rho\partial_-
\rho +2 \,e\, A_-\partial_+\rho
+ b\, e^2\, A_+ A_-\right)
\end{eqnarray}

\noindent where $a$ and $b$ are parameters manifesting regularization,
or equivalently, bosonization ambiguities.  It is simple to verify that
a straightforward application of the usual bosonization rules:
$\bar\psi i\dslash\psi \rightarrow \partial_+\varphi\partial_-\varphi$
and $\bar\psi\gamma_\mu\psi\rightarrow{1\over\sqrt \pi}\epsilon_{\mu\nu}
\partial^\nu\varphi$, which are valid {\it only} when the vector gauge
symmetry is preserved, would just reproduce (\ref{140}) with $a=b=0$.
Subsequently, by functionally integrating out the scalar fields $\varphi$
and $\rho$, exactly yields the two pieces $W[A_+,0]$ and $W[0,A_-]$ given
in (\ref{70}), which is what one obtains by simply putting $A_\pm=0$
directly into the expressions for the vector determinant.  This
reconfirms the invalidity of identifying the chiral determinants
by naively using rules valid for the vector case.

We now show precisely how two independent chiral components (\ref{140}) 
are soldered to yield the LHS of (\ref{60}).  This idea of soldering was 
initially introduced by Stone \cite{MS} and recently exploited by one of 
us \cite{ADW} in a different context.  It consist in lifting the
gauging of a global symmetry to its local version.  Let us then
consider the gauging of the 
following global symmetry of (\ref{140})

\begin{eqnarray}
\label{150}
\delta \varphi &=& \delta\rho=\alpha\nonumber\\
\delta A_{\pm}&=& 0
\end{eqnarray}

\noindent Then it is found from (\ref{140}) that

\begin{eqnarray}
\label{160}
\delta W_+[\varphi] &=& \int d^2x\, \partial_-\alpha \;J_+
(\varphi)\nonumber\\
\delta W_-[\rho]&=& \int d^2x\, \partial_+\alpha \;J_-(\rho)
\end{eqnarray}

\noindent  where,

\begin{equation}
\label{170}
J_\pm(\eta)={1\over{2\pi}}(\partial_\pm\eta +\, e\,A_\pm)\;\;\; ; \;\;\eta=
\varphi , \rho
\end{equation}

\noindent  Next, introduce the soldering field $B_\pm$ so that,

\begin{equation}
\label{180}
W_\pm^{(1)}[\eta] = W_\pm[\eta] -\int d^2x\, B_\mp\, J_\pm(\eta)
\end{equation}

\noindent Then it is easy to verify that the modified action,

\begin{equation}
\label{190}
W[\varphi,\rho]= W_+^{(1)}[\varphi] + W_-^{(1)}[\rho]
 + {1\over{2\pi}} \int d^2x \, B_+ \,B_-
\end{equation}

\noindent is invariant under an extended set of transformations that 
includes (\ref{150}) together with,

\begin{equation}
\delta B_{\pm}= \partial_{\pm}\alpha
\label{191}
\end{equation}

\noindent  Using the equations of motion, the auxiliary soldering
field can be eliminated in favour of the other variables, 

\begin{equation}
\label{200}
B_\pm= 2\pi J_\pm
\end{equation}

\noindent so that the soldered effective
action derived from (\ref{190}) reads,

\begin{equation}
\label{210}
W[\Phi]={1\over {4\pi}}\int d^2x\:\Big{\{}\Big{(}\partial_+
\Phi\partial_-\Phi + 2\,e\, A_+\partial_-\Phi - 2\,e\, A_-
\partial_+\Phi\Big{)} +(a+b-2)\,e^2\,A_+\,A_-\Big{\}}
\end{equation}

\noindent where,

\begin{equation}
\label{220}
\Phi=\varphi - \rho
\end{equation}

\noindent We may now examine the variation of
(\ref{210}) under the lifted gauge transformations,
$\delta\varphi=\delta\rho=\alpha$ and $\delta A_\pm = \partial_\pm\alpha$,
induced by the soldering process. Note that this is just the usual
gauge transformation.
It is easy to see that the expression
in parenthesis is gauge invariant, and by functionally
integrating out the $\Phi$ field one 
verifies that it reproduces (\ref{50}).  Thus,
the soldering process leads to a gauge invariant
structure for $W$ provided

\begin{equation}
\label{230}
a+b-2=0
\end{equation}

\noindent It might appear that there is a whole one parameter class of
solutions.  However Bose symmetry imposes a crucial restriction.  Recall
that in the Feynman graph language this symmetry was an essential
ingredient in preserving compatibility between gauge invariance and
chiral decomposition.  In the soldering process, this symmetry, which
is just the left-right (or $+\, -$) symmetry in (\ref{140}), is preserved
with $a=b$.  Coupled with (\ref{230}) this fixes the parameters to unity
and proves our assertion announced in (\ref{130}).  It may be
observed that the soldering process can be carried through for the
nonabelian theory as well, and a relation analogous to (\ref{210}) is
obtained\footnote{see appendix}.

An alternative way of understanding the fixing of parameters is to recall 
that if a Maxwell term is included in (\ref{130}) to impart dynamics, then 
this corresponds to the chiral Schwinger model\cite{JR}.  It was 
shown that unitarity is violated unless $a$(or $b$)$\geq 1$.  
Imposing (\ref{230}) immediately yields $a=b=1$ as the only valid answer,
showing that the bound 
gets saturated.  It is therefore interesting to note that
(\ref{230}) together with unitarity leads naturally to a Bose
symmetric parametrization. 
In other words, the chiral Schwinger model may have any 
$a\geq 1$, but if two such models with opposite chiralities are soldered to 
yield the vector Schwinger model, then the minimal bound is the unique
choice.  Interestingly, the case $a=1$ implies a massless mode in the
chiral Schwinger model.
The soldering mechanism therefore generates the massive mode of the 
Schwinger model from a fusion of the massless modes in the chiral Schwinger 
models.  

We have therefore explicitly derived expressions for the chiral determinants
(\ref{130}) 
which simultaneously preserve the factorization property (\ref{60}) and 
gauge invariance of the vector determinant.  It was also perceived that the 
naive way of interpreting the chiral determinants as $W[A_+,0]$ or 
$W[0,A_-]$ led to the supposed incompatiblity of factorization with gauge 
invariance since it missed the crossing graphs.  Classically 
these graphs do vanish ($P_+ P_- =0$) so that it becomes evident that this 
incompatibility originates from a lack of properly accounting for the quantum 
effects.  It is possible to interpret this effect, as we will now show,
as a typical quantum mechanical 
interference phenomenon, closely paralleling the analysis in Young's double 
slit experiment.  As a bonus, we provide a new interpretation for 
the Polyakov-Wiegman \cite{PW}
identity. Rewriting (\ref{50}) in Fourier space as

\begin{eqnarray}
\label{240}
W[A_+,A_-] & = & -{N\over 2}\int d^2k\, \{A^*_+(k){k_-\over k_+}A_+(k)
+ A^*_-(k)
{k_+\over k_-}A_-(k) - 2A^*_+(k)A_-(k)\} \nonumber\\
&=& -{N\over 2}\int d^2k\, \mid\sqrt{k_-\over k_+}A_+(k) -
\sqrt{k_+\over k_-}A_-(k)
\mid^2
\end{eqnarray}

\noindent immediately displays the typical quantum mechanical interference 
phenomenon, in close analogy to the optical example,

\begin{equation}
\label{250}
W[A_+,A_-]= -{N\over 2}\int d^2k\,\left(\mid\psi_+(k)\mid^2
+\mid\psi_-(k)\mid^2 
+ 2\cos\theta \psi^*_+(k)\psi_-(k)\right)
\end{equation}

\noindent with $\psi_\pm (k)=\sqrt{k_\mp\over k_\pm}A_\pm(k)$ 
and $\theta=\pm\pi$, simulating the roles of the amplitude and the phase,
respectively.  Note that in one space dimension, these are 
the only possible values for the phase angle $\theta$ between the 
left and the right movers. The dynamically  
generated mass arises from the interference 
between these movers, thereby preserving gauge invariance. Setting either 
$A_+$ or $A_-$ to vanish, destroys 
the quantum effect, very much like closing one slit in the optical 
experiment destroys the interfernce pattern.   
Although this analysis was done for the abelian theory, it is
straitghtforward to perceive that the effective action for a
nonabelian theory can also be expressed in the form of an absolute
square (\ref{240}), except that there will be a repetition of
copies depending on the group index.
This happens because only the two-legs graph has
an ultraviolet divergence, leading to the interference (mass) term.
The higher legs graphs are all finite, and satisfy the naive
factorization property.

It is now simple to see that (\ref{240}) represents an abelianized
version of the Polyakov 
Wiegman identity by making a familiar change of variables, 

\begin{eqnarray}
\label{260}
A_+&=&{i\over e}U^{-1}\partial_+ U\nonumber\\
A_-&=&{i\over e}V\partial_- V^{-1}
\end{eqnarray}

\noindent where, in the abelian case, the matrices $U$ and $V$ are given as,

\begin{equation}
\label{270}
U=\exp\{i\varphi\}\;\;\;\; ; \;\;\;\; V=\exp\{-i\rho\}\;\;\;\; ; \;\;\;\; 
UV =\exp\{i\Phi\}
\end{equation}

\noindent  with $\Phi$ being the gauge invariant soldered field
introduced in (\ref{220}).  It is possible to recast (\ref{240}),
in the coordinate space,  as

\begin{equation}
\label{280}
W[UV]=W[U] + W[V] + {1\over{2\pi}}\int d^2x\, \left(U^{-1}\partial_+ U\right)
\left(V\partial_- V^{-1}\right)
\end{equation}

\noindent which is the Polyakov-Wiegman identity, satisfying gauge invariance.
The result can be extended to the nonabelian case since, as already mentioned,
the nontrivial interference term originates from the two-legs graph which
has been taken into account. It is now relevant to point out that the important crossing piece in
either (\ref{240}) or (\ref{280}) is conventionally \cite{PW, AAR}
interpreted as a contact (mass) term, or a counterterm, necessary to
restore gauge invariance. In our analysis, on the contrary, this term
was uniquely specified from the interference between the left and right
movers in one space dimension, automatically providing gauge invariance.
This is an important point of distinction.

To conclude, our analysis clearly revealed that no obstruction
to gauge invariance is posed by the chiral decomposition of the 2D
fermionic determinant. 
The claimed obstruction actually results from an incorrect interpretation 
of the chiral determinants. Bose symmetry gave a precise way of making sense 
of these determinants which were explicitly computed by considering the dual 
descriptions of decomposition
as well as soldering. The close interplay between Bose symmetry and
gauge invariance was illustrated in both these ways of looking at the
fermionic determinant. At the dynamical level it was also shown how this
symmetry is instrumental in fusing the massless modes of the left and right
chiral Schwinger models to yield
the single massive mode of the vector Schwinger model. Indeed it was
explicitly shown that this mass generation is the quantum interference effect
between the two chiralities, closely resembling the corresponding effect
in the double slit optical experiment. This led us to provide a novel
interpretation of the Polyakov-Wiegman identity.

Our analysis indicated that the $a=1$ regularisation for
the determinant of the
Chiral Schwinger Model was important leading to interesting effects. 
This parametrisation was also found to be useful in a different context
\cite{W}. On the other hand 
much of the usual analyses is confined to the 
$a=2$ sector \cite{L}.

Finally, to put this work in a proper perspective it may be useful to once
again remind the importance of Bose symmetry. It is an essential
ingredient in getting the classic ABJ anomaly from the triangle
graph \cite{R, SA}. Just imposing gauge invariance on the
vector vertices does not yield the cherished result. Bose symmetry coupled
with gauge invariance does the job. This symmetry also played a crucial
role in providing a unique structure for the 1-cocycle that is mandatory
for smooth bosonization\cite{RB, DN}.  It is therefore not surprising that 
Bose symmetry provided the definite guideline in preserving the compatibility
between gauge invariance and chiral decomposition or soldering.\vspace{0.3cm}\\

\noindent Acknowledgments.  One of the authors (RB) thanks
the members of the Department of Physics of UFRJ for their
kind hospitality and CNPq for providing financial support.  The other
authors are partially supported by CNPq, CAPES, FINEP and FUJB , Brasil.

\newpage
\noindent {\bf APPENDIX}
\bigskip
\bigskip

Here we explicitly show the soldering mechanism in the nonabelian context.
The expressions for the chiral determinants analogous to (\ref{140}) are
given by \cite{LR},
\begin{eqnarray}
W_+[g]&=& I_{wzw}^{(-)}[g] -\frac{ie}{2\pi}\int d^2x 
tr(A_+g^{-1}\partial_-g)-\frac{e^2a}{4\pi}\int d^2x tr(A_+A_-)\nonumber\\
W_-[h]&=& I_{wzw}^{(+)}[h] -\frac{ie}{2\pi}\int d^2x 
tr(A_-h^{-1}\partial_+h)-\frac{e^2b}{4\pi}\int d^2x tr(A_+A_-)
\label{a10}
\end{eqnarray}
where the Wess-Zumino-Witten functional at the critical point $(n=\pm 1)$
is given by (for details and the original papers, see
\cite{AAR}),
\begin{equation}
I_{wzw}^\pm[k]=\frac{1}{4\pi}\int d^2x tr(\partial_+k\partial_-k^{-1})
\mp\frac{1}{12\pi}\Gamma_{wz}[k]\,\,\,\,\, ;k=g, h
\label{a11}
\end{equation}
with the familiar Wess-Zumino term defined over a 3D manifold with the 
two-dimensional Minkowski space-time as its boundary,
\begin{equation}
\Gamma_{wz}[k]=\int d^3x \epsilon^{lmn}\,tr(k^{-1}\partial_l k\, 
k^{-1}\partial_m k\, k^{-1}\partial_n k)
\label{a12}
\end{equation}
In the above equations $g$ and $h$ are the elements of some compact Lie group
and the parameters $a$ and $b$ manifest the regularisation or bosonisation
ambiguities. Let us next consider the gauging of the global right and left
chiral symmetries analogous
to (\ref{150}),
\begin{eqnarray}
\delta g&=&\omega g\nonumber\\
\delta h&=&h\omega\nonumber\\
\delta A_{\pm}&=&0
\label{a13}
\end{eqnarray}
where $\omega$ is an infinitesimal element of the algebra of the corresponding
group. Note the order of $\omega$ which occurs once from the left and once
from the right to properly account for the two chiralities. In the abelian
example, this just commutes and the ordering is unimportant leading to a
unique transformation in (\ref{150}). Under (\ref{a13}), the relevant
variations are found to be,
\begin{eqnarray}
\delta W_+[g]&=&\int d^2x tr(\partial_-\omega J_+(g))\nonumber\\
\delta W_-[h]&=&\int d^2x tr(\partial_+\omega J_-(h))
\label{a14}
\end{eqnarray}
where,
\begin{equation}
J_\pm=\frac{-1}{2\pi}\Big(\partial_\pm kk^{-1}+iekA_\pm k^{-1}\Big)
\label{a15}
\end{equation}
Now introduce the soldering field $B_\pm$ which transforms as,
\begin{equation}
\delta B_\pm=\partial_\pm\omega -[B_\pm, \omega]
\label{a16}
\end{equation}
whose abelian version just corresponds to (\ref{191}). Then it may be
checked that the following effective action,
\begin{equation}
W[g, h]=W_+[g]+W_-[h] -\int d^2x tr\Big(B_-J_+(g)+B_+J_-(h)
+\frac{1}{2\pi}B_+B_-\Big)
\label{a17}
\end{equation}
is invariant under the complete set of transformations. The auxiliary 
soldering field is eliminated, as usual, in favour of the other variables,
by using the equations of motion,
\begin{equation}
B_\pm=-2\pi J_\pm(k)
\label{a18}
\end{equation}
The soldered effective action directly follows from (\ref{a17}) on substituting
this solution,
\begin{eqnarray}
W[G]&=&I_{wzw}^+[G]+\frac{ie}{2\pi}\int d^2xtr\Big(A_-G^{-1}\partial_+G
-A_+\partial_-GG^{-1}\Big)\nonumber\\
&-&\frac{e^2}{2\pi}\int d^2x tr\Big(A_+GA_-G^{-1}-\frac{a+b}{2}A_+A_-\Big)
\label{a19}
\end{eqnarray}
where $G=g^{-1}h$. Once again gauge invariance under the conventional set
of transformations in which $A_\pm$ changes as a potential, is recovered
only if $a+b=2$, exactly as happened in the abelian case. Imposing Bose
symmetry leads to the unique choice $a=b=1$, completely determining the
structure for the separate chiral components. 
Incidentally, by including the Yang-Mills term to impart dynamics so that these
models become chiral $QCD_2$, it was found that unitarity could be preserved
only for $a, b\geq 1$ \cite{LR}. Coupled 
with the above noted restriction, this leads
to the Bose symmetric parametrisation. 
It is easy to see that
(\ref{a19}) reduces to the abelian result (\ref{210}) by
setting $G=\exp(i\Phi)$. Observe that the soldering
was done among the chiral components having opposite critical points. Any
other combination would fail to reproduce the gauge invariant result. 
Indeed the gauge invariant effective action, being a functional of 
$G=g^{-1}h$, can be obtained
by soldering effective actions (which are functionals of $g$ and $h$)
with opposite criticalities since changing $g\rightarrow g^{-1}$ converts
the Wess-Zumino-Witten functional from one criticality to the other.
The relevance of opposite criticality was also noted in another context
involving smooth nonabelian bosonisation \cite{DN1}.

This nonabelian exercise, however, clearly reveals 
that the physics of the problem
of chiral soldering (or decomposition) and the role of Bose symmetry
is contained in the abelian sector. The rest is a matter of 
technical detail. Indeed, following similar steps, it is also
possible to discuss chiral decomposition for the nonabelian case and obtain
identical conclusions.

\end{document}